# Parameter Extension Simulation of Turbulent Flows

Yan Jin

Center of Applied Space Technology and Microgravity (ZARM), University of Bremen, 28359, Bremen, Germany, e-mail: yan.jin@zarm.uni-bremen.de

**Abstract**

Parameter extension simulation (PES) as a mathematical method for simulating turbulent flows has been proposed in the study. It is defined as a calculation of the turbulent flow for the desired parameter values with the help of a reference solution. A typical PES calculation is composed of three consecutive steps: Set up the asymptotic relationship between the desired solution and the reference solution; Calculate the reference solution and the necessary asymptotic coefficients; Extend the reference solution to the desired parameter values. A controlled eddy simulation (CES) method has been developed to calculate the reference solution and the asymptotic coefficients. The CES method is a special type of large eddy simulation (LES) method in which a weight coefficient and an artificial force distribution are used to model part of the turbulent motions. The artificial force distribution is modeled based on the eddy viscosity assumption. The reference weight coefficient and the asymptotic coefficients can be determined through a weight coefficient convergence study. The proposed PES/CES method has been used to simulate four types of turbulent flows. They are decaying homogeneous and isotropic turbulence, smooth wall channel flows, rough wall channel flows, and compressor blade cascade flows. The numerical results show that the 0-order PES solution (or the reference CES solution) has a similar accuracy as a traditional LES solution, while its computational cost is much lower. A higher order PES method has an even higher model accuracy.

**Keywords**: Turbulence modeling; computational fluid dynamics; eddy viscosity assumption; large eddy simulation.

## 1 Introduction

Turbulent flows are fluid motions which are characterized by chaotic changes in pressure and flow velocity. They are usually accompanied with three-dimensional and transient velocities, large range of time and length scales, and nonlinear pressure gradients [1]. Turbulent flows are often observed phenomena in everyday surroundings and prevalent processes in industry. Intensive studies with respect to predicting turbulent flows have been carried out in recent 100 years. However, turbulence is a process with high complexity and the physics of turbulence is still not fully understood. Due to this reason, prediction of turbulent flows is extremely difficult.

The most often used CFD approaches for solving turbulent flows include direct numerical simulation (DNS) [2, 3], Reynolds averaged Navier-Stokes equations simulation (RANS) [4-6], large eddy simulation (LES) [7-9], scale-adaptive simulation (SAS) [10], and detached eddy simulation (DES) [11]. The Navier-Stokes equations are numerically solved without any turbulence model in DNS. It is the most accurate CFD approach whereas it is rarely employed in industrial applications due to its high computational costs, especially for high Reynolds number problems. Practical solution of a turbulent flow more or less requires a turbulence model.

RANS is an often used method for solving turbulent flows in industry due to its low computational costs and reasonable results. Many RANS models were developed based on the eddy viscosity assumption proposed by Boussinesq [12]. RANS methods have been widely accepted for solving turbulent flows in engineering applications. However, the assumptions used in RANS may sometimes lead to considerable model errors and uncertainties in CFD results.

LES is a compromise between DNS and RANS, in which the large eddies are solved directly like in DNS while the subgrid-scale stresses (SGS) are modeled. LES is more accurate than RANS, however, it is still too expensive for many engineering problems which have a high Reynolds number. Some other turbulence models blend LES and RANS; the examples are the Scale-Adaptive simulation (SAS) [10] and Detached Eddy Simulation (DES) [11]. Since the deficiencies of LES and RANS also apply to these approaches, they still need further improvement and validations.

The purpose of the present study is developing a new method for calculating turbulent. It is developed based on the asymptotic theory. In an early paper, the asymptotic method has been suggested by Carey & Mollendorf [13] for studying viscosity effects for natural convection flows. Later, Herwig and his colleagues further extended the method and performed comprehensive studies with respect to the variable property effects on flow and heat transfer problems, see [14-19]. The temperature dependent fluid properties under consideration include the density, viscosity, heat capacity, and thermal conductivity. A Taylor series expansion of all properties with respect to temperature has been made in this asymptotic method. The asymptotic coefficients (A-coefficients) were introduced to calculate the solutions for the variable property fluids [15].

Later, Bünger & Herwig [20] introduced a higher order coefficient method (HOC) to calculate the A-coefficients in the asymptotic method. When applied in a more general sense, the HOC approach is called asymptotic computational fluid dynamics (ACFD) since CFD solutions are used to calculate the certain A-coefficients in asymptotic expansions. Jin & Herwig [21] demonstrated how to account for the variable property effects on the mixed convection with the ACFD method. A more efficient way for calculating the asymptotic coefficients (or A-coefficients) in the Taylor series expansion were suggested in [22].

Jin & Herwig [23] summarized the previous studies and proposed the parameter extension method (PEM) for calculating flow and heat transfer problems. Up to now, the PES is still not a flow simulation method but a method for accounting for variable property effects. A complex turbulent flow, particularly at a high Reynolds number, still cannot be simulated using the PEM. One reason is that it is difficult to determine the reference solution which is qualitatively similar to the desired solution but less expensive. In addition, calculation of the A-coefficients using the HOC method is computationally expensive. The errors for the calculated A-coefficients and the neglected high order terms may have significant effects on the accuracy of the PEM results.

In the current study, based on the PEM, we will propose a turbulent flow simulation method, i.e., parameter extension simulation (PES). The structure of the paper is as follows: The PES method is introduced in section 2. The numerical methods for a PES solution is introduced in section 3. Four types of turbulent flows are solved by using PES in section 4 to demonstrate its applications. Finally, the conclusions are given in section 5.

## 2. Parameter extension simulation

As a mathematical method for simulating turbulent flows, parameter extension simulation (PES) is defined as a calculation of the turbulent flow for the desired parameter values with the help of a solution for the initial parameter values, i.e., the reference solution. The basic idea for PES is that often a continuous change of parameter values will lead to a corresponding continuous change in the solution. Mathematically the PES method corresponds to a regular perturbation approach with the (small) deviations of the parameters from their initially defined reference values as (small) perturbation parameters. Thus, the desired solution $R$ can be extended from the reference solution $R_0$ with the help of the asymptotic expansion. In order to end up with an asymptotic approximation of a solution three consecutive steps are necessary. They are:

**STEP 1**: Set up the asymptotic relationship between the desired solution and the reference solution.

**STEP 2**: Calculate the reference solution $R_0$ and the necessary A-coefficients.

**STEP 3**: Extend $R_0$ to the desired parameter values.

A PES method is proposed in this section. The governing equations for calculating the reference solution and the A-coefficients are introduced in sub-section 2.1. The sub-models for closing the governing equations are introduced in sub-sections 2.2 and 2.3.

2.1. Governing equations

The reference solution should be qualitatively similar to the desired solution which is usually a DNS solution. In addition, it should be less expensive. We may use a LES solution as the reference solution since it approaches to the DNS solution as the mesh size $\Delta$ is reduced to zero. However, the mesh is often non-uniformly distributed for a complex flow problem, thus it is difficult to perform an asymptotic study with respect to $\Delta$.

Hereby we propose a special type of LES for calculating the reference solution. The governing equations are the filtered Navier-Stokes equations. For an incompressible flow, they read

$$\frac{\partial \tilde{u}_i}{\partial x_i} = 0 \tag{2.1}$$

$$\frac{\partial \tilde{u}_i}{\partial t} + \frac{\partial (\tilde{u}_i \tilde{u}_j)}{\partial x_j} = -\frac{\partial \tilde{p}}{\partial x_i} + \nu \frac{\partial^2 \tilde{u}_i}{\partial x_j^2} + \phi F_i + g_i \tag{2.2}$$

The tilde ˜ denotes a filtered variable in the model equations. $g_i$ is a body force. An artificial force $\phi F_i$ is introduced into the momentum equation in order to model part of the turbulent motions. $F_i$ determines the distribution of the artificial force, while $\phi$ is a weight coefficient which determines the strength of the artificial force. Since the eddies to be directly resolved can be controlled by adjusting the $\phi$ value, this type of simulation is called controlled eddy simulation (CES). We solve Eqs. (2.1)-(2.2) for $\phi = \phi_0$ to obtain the reference solution.

CES is a special type of LES. However, the artificial force for the CES is independent of the mesh size $\Delta$. Instead, a CES solution approaches to a DNS solution as the weight coefficient $\phi \to 0$. A small perturbation parameter $\varepsilon_\phi$ can be defined as

$$\varepsilon_\phi = \phi_0 - \phi \tag{2.3}$$

An asymptotic expansion can be made from the reference solution $R_R(\varepsilon_\phi = 0)$:

$$R(\varepsilon_\phi) = R_0 + A_1 \varepsilon_\phi + A_2 \varepsilon_\phi^2 + \cdots A_{n-1} \varepsilon_\phi^{n-1} + O(\varepsilon_\phi^n) \tag{2.4}$$

where $A_i, i = 1 \cdots n - 1$ are the asymptotic coefficients. The DNS solution $R_D$ can be obtained if we set $\varepsilon_\phi$ to the value of $\phi_0$, i.e.,

$$R_D = R_0 + A_1\phi_0 + A_2\phi_0^2 + \cdots A_{n-1}\phi_0^{n-1} + O(\phi_0^n) \tag{2.5}$$

We may use the reference CES solution $R_0$ to approximate the DNS solution $R_D$, leading to a 0-order PES solution (or reference CES solution). We are able to determine the A-coefficients with more CES solutions for different $\phi$ values, leading to a higher order PES method.

*2.2. Modified mixing length (ML+) model*

A modified mixing length (ML+) model was developed to determine the artificial force distribution $F_i$. The eddy viscosity hypothesis is adopted in this model. $F_i$ is calculated as

$$F_i = \frac{\partial}{\partial x_j}\left(2\nu_{eff}\tilde{s}_{ij}\right) \tag{2.6}$$

where $\nu_{eff}$ is an effective eddy viscosity. Similar to the classic mixing length model [24], the eddy viscosity $\nu_{eff}$ is determined by the product of a mixing length $l_{mix}$ and a characteristic velocity $\vartheta_{mix}$. The characteristic velocity $\vartheta_m$ is calculated as

$$\vartheta_{mix} = l_{mix}|\tilde{s}_{ij}| \tag{2.7}$$

where $\tilde{s}_{ij} = \frac{1}{2}\left(\frac{\partial \tilde{s}_{ij}}{\partial x_j} + \frac{\partial \tilde{s}_{ij}}{\partial x_i}\right)$ is the stain rate of the filtered flow field. $|\tilde{s}_{ij}| = \left(2\tilde{s}_{ij}\tilde{s}_{ij}\right)^{1/2}$ is the magnitude of $\tilde{s}_{ij}$.

van Driest [25] suggested that $l_m$ can be calculated as

$$l_{mix} = \kappa y_w F_D(y_w^+) \tag{2.8}$$

where $\kappa = 0.41$ is the von Kármán constant, $y_w$ is the distance from the wall. The damping function $F_D(y_w^+)$ is calculated as

$$F_D(y_w^+) = 1 - exp(-y_w^+/A^+) \tag{2.9}$$

where $y_w^+ = \frac{y_w u_\tau}{\nu}$ is the dimensionless distance from the wall. $u_\tau$ is the friction velocity. $A^+ = 26$ is a model constant.

Since it is difficult to calculate $y_w^+$ for a complex geometry, $F_D(y_w^+)$ in Eq. (2.8) is replaced with a function of $y_s^+$, which is defined as

$$y_s^+ = \sqrt{\frac{|\partial k/\partial t|}{\nu|s_{ij}|^2}} \tag{2.10}$$

where $k = \frac{1}{2}u_k^2$ is the instantaneous kinetic energy. $|\partial k/\partial t|$ characterizes the strength of the transient fluctuation. It may be noted that $y_s^+$ becomes zero if the flow is steady. Therefore, the model also applies to steady laminar flows, for which the artificial force $F_i$ is zero. The modified effective viscosity is calculated as,

$$\nu_{eff} = l_{mix}'^2 |\tilde{s}_{ij}| \quad (2.11)$$

The square of the modified mixing length $l_{mix}'^2$ is calculated as

$$l_{mix}'^2 = \min\left\{\kappa^2 y_w^2 \left[1 - \exp\left(-\frac{y_s^+}{B^+}\right)\right]^n, l_c^2\right\} \quad (2.12)$$

where $l_c$ is the characteristic length of the flow regime. If the flow is not bounded by walls, the mixing length is equivalent to the characteristic length $l_c$. Here we use $B^+ = 1$ and $n = 1$ as the model constants.

## 2.3. Reference weight coefficient $\phi_0$ and A-coefficients

Similar to the mesh size $\Delta$ for a traditional LES, the weight coefficient $\phi$ determines the cutoff eddy size for a CES solution. More turbulent motions will be modeled as the value of $\phi$ increases, thus the computational costs can be reduced. However, the increase of $\phi$ may result in a higher model error. Therefore, the value of $\phi_0$ for calculating the reference solution should be determined as a compromise between the model accuracy and the computational cost.

In a LES method, it is often assumed that all kinetic energy transported from the filtered flow field to the unresolved small eddies will be eventually dissipated without affecting the filtered flow field. This assumption may be better interpreted using the integral equation for the mechanical energy. Multiplying Eq. (2.2) with $\tilde{u}_i$, averaging it over time, and integrating it in the whole flow domain, we may obtain the integral equation for the mechanical energy,

$$\tilde{L} = -\int_A \left(\overline{\tilde{u}_i \tilde{p}^*} - 2\overline{(\nu + \nu_{eff})\tilde{s}_{ij}\tilde{u}_j}\right) n_i dA' + gu_m = \int_V \bar{\tilde{\varepsilon}} dv \quad (2.13)$$

where $\tilde{L}$ denotes the overall loss of the filtered mechanical energy. $u_m$ is the mean velocity. It is composed of the change of the filtered stagnation pressure $-\int_A \overline{\tilde{u}_i \tilde{p}^*} n_i dA'$, the work done by the viscous force at the boundary surfaces $\int_A 2\overline{(\nu + \nu_{eff})\tilde{s}_{ij}\tilde{u}_j} n_i dA'$, and the work done by the body force $gu_m$. $\tilde{p}^* = \frac{1}{2}\tilde{u}_k^2 + \tilde{p}$ is the filtered stagnation pressure. The pseudo dissipation rate for the filtered flow field $\tilde{\varepsilon}$ is composed of the directly resolved dissipation rate $\tilde{\varepsilon}_d$ and the modeled dissipation rate $\tilde{\varepsilon}_m$. They are defined as

$$\tilde{\varepsilon} = \tilde{\varepsilon}_d + \tilde{\varepsilon}_m; \tilde{\varepsilon}_d = 2\nu\tilde{s}_{ij}\tilde{s}_{ij}; \tilde{\varepsilon}_m = 2\nu_{eff}\tilde{s}_{ij}\tilde{s}_{ij} \qquad (2.14)$$

$\tilde{\varepsilon}_m$ indicates the kinetic energy which is transported from the filtered flow field to the unresolved eddies which are smaller than the cutoff eddy size. If $\phi$ is set to 0, Eq. (2.13) becomes

$$L = -\int_A \left(\overline{u_\iota p^*} - 2\overline{\nu s_{\iota j} u_j}\right) n_i dA' + gu_m = \int_V \bar{\varepsilon} dv \qquad (2.15)$$

where $\bar{\varepsilon}$ is the local mean dissipation rate obtained from a DNS solution.

According to the eddy-cutoff assumption mentioned above, the kinetic energy which is transported to the unresolved eddies will be all dissipated if the cutoff eddy size is sufficiently small. Therefore, $\int_V (\bar{\tilde{\varepsilon}}_d + \bar{\tilde{\varepsilon}}_m)dv)$ will be equal to $\int_V \bar{\varepsilon}dv)$ if $\phi$ is smaller than a critical value $\phi_c$. Thus, a $\phi$-convergence study with respect to $\int_V (\bar{\tilde{\varepsilon}}_d + \bar{\tilde{\varepsilon}}_m)dv)$ (or $\tilde{L}$) may be performed to determine whether the reference solution is qualitatively similar to the DNS solution. Similar to a mesh-convergence study, CES solutions for at least another two $\phi$ values, $\hat{t}\phi_0$ and $\hat{t}^2\phi_0$, are needed for a $\phi$-convergence study. $\hat{t}$ is a factor which is larger than 1. The 0- and 1- order PES solutions can be calculated as:

$$R_0 = R(\phi_0) \qquad (2.16)$$

$$R_1 = R(\phi_0) + A_1\phi_0; A_1 = \frac{R(\phi_0) - R(\hat{t}\phi_0)}{(\hat{t} - 1)\phi_0} \qquad (2.17)$$

A higher order PES can be calculated using the Richardson extrapolation: The Richardson-PES solution is calculated as

$$R_{Ri} = R(\phi_0) + \frac{R(\phi_0) - R(\hat{t}\phi_0)}{\hat{t}^{\hat{p}} - 1} \qquad (2.18)$$

where the observed order of accuracy $\hat{p}$ is calculated as:

$$\hat{p} = -ln\left(\frac{R(\phi_0) - R(\hat{t}\phi_0)}{R(\hat{t}\phi_0) - R(\hat{t}^2\phi_0)}\right)/ln(\hat{t}) \qquad (2.19)$$

It may be noticed that a full asymptotic expansion is approximated in Eq. (2.18). The error for the reference solution $\delta_{R0}$ can be estimated by

$$\delta_{R0} = \frac{|R_{Ri} - R_0|}{R_0} \qquad (2.20)$$

## 3. Numerical methods

Eqs. (2.1)-(2.2) can be solved using a finite volume method (FVM). The solver was developed based on the open source CFD code OpenFoam 16.06+. The solutions are advanced in time with the second order implicit backward method. To compute the derivatives of the velocity, the variables at the interfaces of the grid cells are obtained with linear interpolation. With the solutions at the interfaces, a second order central difference scheme can be gained for spatial discretization.

The pressure at the new time level is determined by the Poisson equation. The velocity is corrected by the Pressure-Implicit with Splitting of Operators (PISO) pressure-velocity coupling scheme. The solver has received intensive validations in our previous studies, see [26-29] as examples.

In [28], we have introduced an accuracy measure $\delta_{num}$ for assessing the numerical error of a numerical solution. $\delta_{num}$ is defined as

$$\delta_{num} = \frac{\tilde{L} - \int_V \bar{\tilde{\varepsilon}} dv}{\tilde{L}} \tag{3.1}$$

The value of $\delta_{num}$ for an ideal numerical solution is zero. For a real numerical solution, however, due to the existence of the numerical dissipation, $\delta$ is usually larger than 0. For example, the $\delta_{num}$ values are 0.1%-1.5% for the DNS results for a flow in a channel with smooth walls (see [30]). They are 7%-10% for more complicated turbulent flows such as a flow in a channel with rough walls (see [26] and [27]) or a turbulent flow in porous media (see [28] and [29]).

## 4. Test cases of application

We demonstrate the application of PES by four cases. They are decaying homogeneous and isotropic turbulence, smooth wall channel flows, rough wall channel flows, compressor blade cascade flows.

4.1. Decaying homogeneous and isotropic turbulence

The first test case is the decaying homogeneous and isotropic turbulence in a box with the size $2\pi \times 2\pi \times 2\pi$. Periodic boundary conditions are given in all three directions. The Reynolds number based on the non-dimensional viscosity $\text{Re} = 1/\nu$ is $10^5$. In the initial field, the velocity amplitude $Ea$ for the wave number $\mathbf{k} = (k_1, k_2, k_3)$ is given as

$$Ek(\mathbf{k}) = Ea \frac{|\mathbf{k}|^4}{k_0} \exp\left[-2\left(\frac{|\mathbf{k}|}{k_0}\right)^2\right] \tag{4.1}$$

where $Ea = 10$ and $k_0 = 5$ are constants.

The CES solution for $\phi = 2 \times 10^{-5}$ was used as the reference solution. Another two CES solutions for $\phi = 4 \times 10^{-5}$ and $8 \times 10^{-5}$ were used to perform the $\phi$-convergence study. The ratio between the modeled losses magnitude and the overall losses magnitude $\delta_m$ is shown in Fig. 1. It is calculated as

$$\delta_m = \int_V 2\nu_{eff} \tilde{s}_{ij} \tilde{s}_{ij} dv \Big/ \int_V 2(\nu_{eff} + \nu) \tilde{s}_{ij} \tilde{s}_{ij} dv \tag{4.2}$$

It can be seen in Fig. 1 that $\delta_m$ increases with $\phi$.

The observed accuracy $\hat{p}$ was calculated using Eq. (2.19), where $\hat{t}$ is equal to 2. $R = \frac{1}{T}\int_T k dt$ is the averaged kinetic energy during the time under consideration. The calculated observed accuracy $\hat{p}$ is 0.69. The 0-order, 1-order, and Richardson- PES solutions are compared with the DNS results in Fig. 2. It may be noticed that the higher order PES solutions are more accurate than the 0-order PES solution.

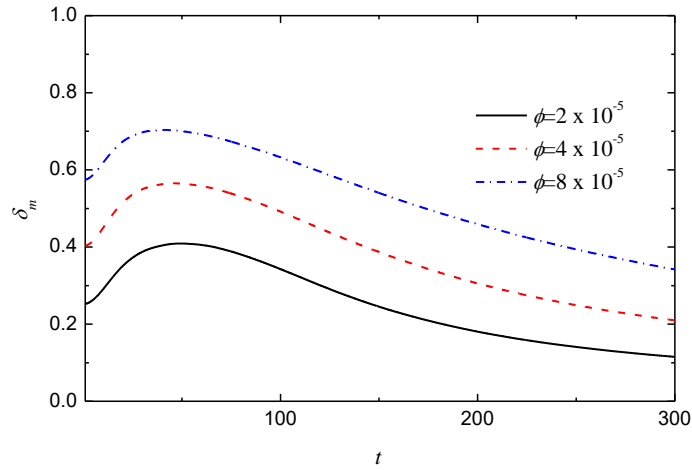

Figure 1. History for the fraction of the modeled losses.

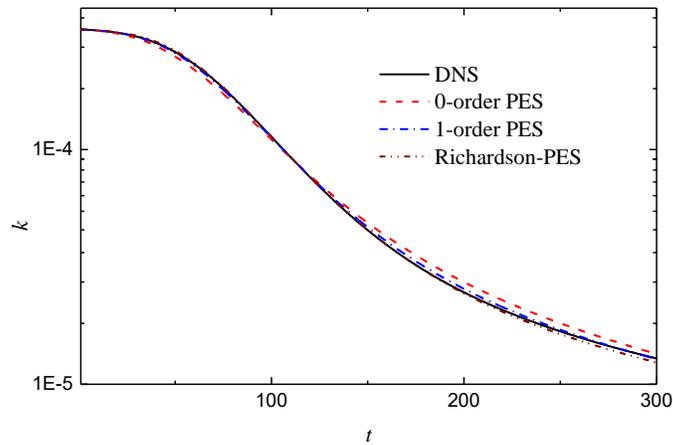

Figure 2. History for the kinetic energy.

Fig. 3 shows the relationship between the kinetic energy and the observed decay rate $dk/dt$. The 0-order PES result (reference CES solution) is compared with the DNS solution. The two solutions are close to each other and the predicted $dk/dt \sim k$ scaling is close to $dk/dt \sim k^{3/2}$, which is the scaling law for an inertial-range spectrum, see[1].

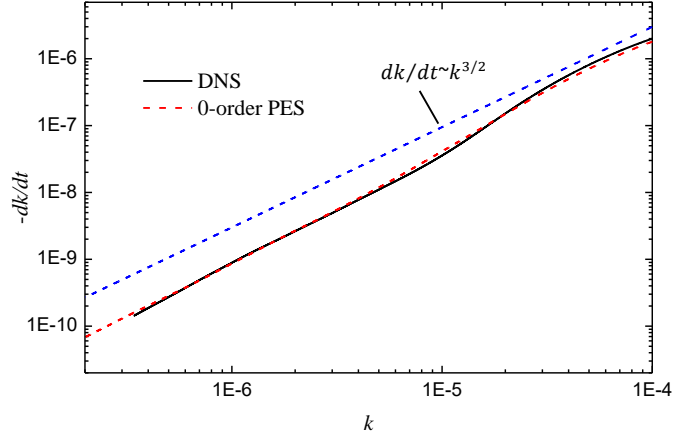

Figure 3. Kinetic energy versus the observed decay rate.

4.2. Smooth wall channel flows

The geometry of the test case is a channel bounded with two smooth walls. The Reynolds number for this type of flows is defined as

$$\text{Re} = \frac{u_m H}{\nu} \quad (4.3)$$

where $u_m$ is the mean velocity and $H$ is the half channel height. The computational domain size is $6.28H \times 2H \times 3.14H$. The turbulent flows for different Reynolds numbers were calculated using PES and traditional LES methods. The subgrid models used in the traditional LES solutions include the Smagorinsky model [7], the k-equation transport model [8], and the WALE model [9]. All test cases were calculated with the same mesh resolution ($225 \times 160 \times 225$). Another higher-resolution mesh resolution ($256 \times 212 \times 256$) was used to carry out the mesh independence study.

For a channel flow, the magnitude of the overall losses $\tilde{L}$ corresponds to the friction coefficient $f$ which is defined as

$$f = \frac{8gH}{u_m^2} \quad (4.4)$$

where $g$ is the magnitude of the body force. The $f$-$\phi$ convergence study is shown in Fig. 4. The results for $\phi = 0.003$ are used as the reference solution. Fig. 4 shows that $f$ decreases almost linearly as $\phi$ increases from 0.003 to 0.005, thus the observed accuracy $\hat{p}$ is about 1. Therefore, the Richardson-PES solution $f_{Ri}$ is approximately equivalent to 1-order PES solution $f_1$.

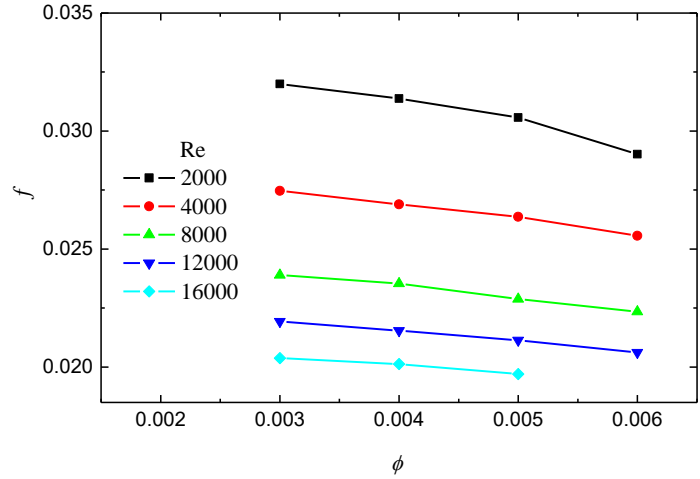

Figure 4. Friction coefficient $f$ versus weight coefficient $\phi$.

Fig. 5 shows the PES and traditional LES results of $f$ for different Reynolds numbers. The numerical results were compared with the DNS results by Kis [30]. The 0-order PES solutions are similar to the WALE model results at high Reynolds numbers (Re $\geq$ 12000). Both of them are more accurate than the Smagorinsky model and k-equation model results. The accuracy for the PES solution can be further improved when it is corrected using a higher order expansion. The 1-order (or Richardson-) PES solution has the highest accuracy among all the solutions under consideration.

Figure 6 shows the numerical error $\delta_{num}$ defined as Eq. (3.1) for the PES and traditional LES solutions. It can be seen that the 0-order PES (or reference CES) solution has a smaller numerical error than the other LES solutions. Since $\delta_{num}$ decreases as the mesh resolution is improved, to achieve the same value of $\delta_{num}$, we expect that a PES solution requires fewer mesh cells than the other LES solutions.

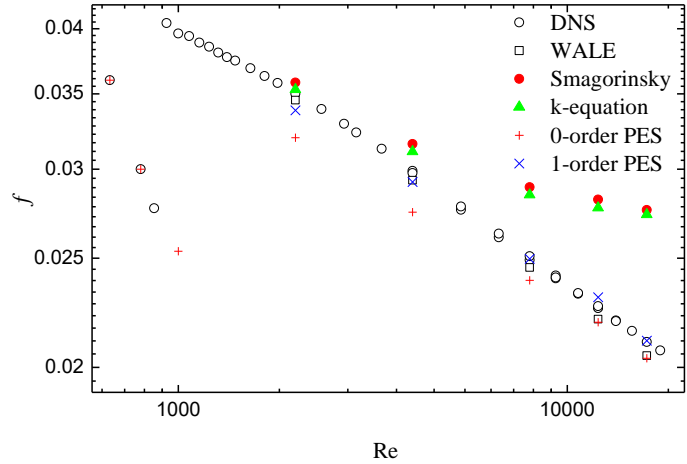

Figure 5. Friction coefficient $f$ versus Reynolds number Re.

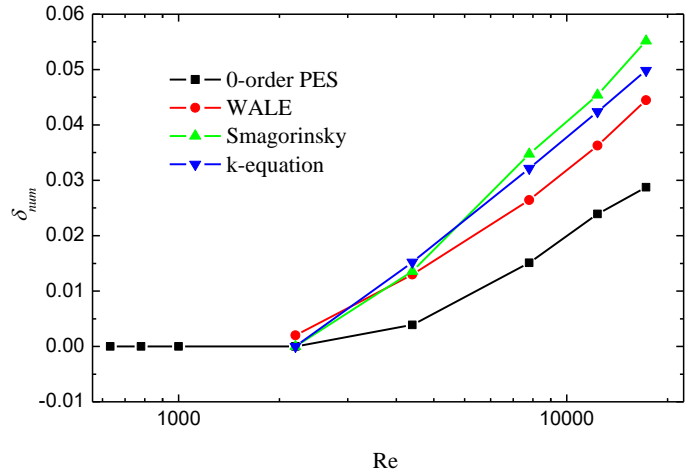

Figure 6. Numerical error $\delta_{num}$ versus Reynolds number Re, Mesh resolution: $225 \times 160 \times 225$. The 0-order PES (reference CES) results are compared with the traditional LES results.

Figure 7 shows the PES and DNS results for the mean velocity $\bar{u}_1$ and the normal Reynolds stresses $\overline{u'_i u'_i}$. The 0-order PES (reference CES) results are in reasonable accordance with the DNS results. The results of $\bar{u}_1$ and $\overline{u'_i u'_i}$ are not as sensitive to $\phi$ as the friction coefficient $f$. It can be seen that the 1-order PES solutions for $\bar{u}_1$, $\overline{u'_2 u'_2}$, and $\overline{u'_3 u'_3}$ are only mildly different from the 0-order PES solutions.

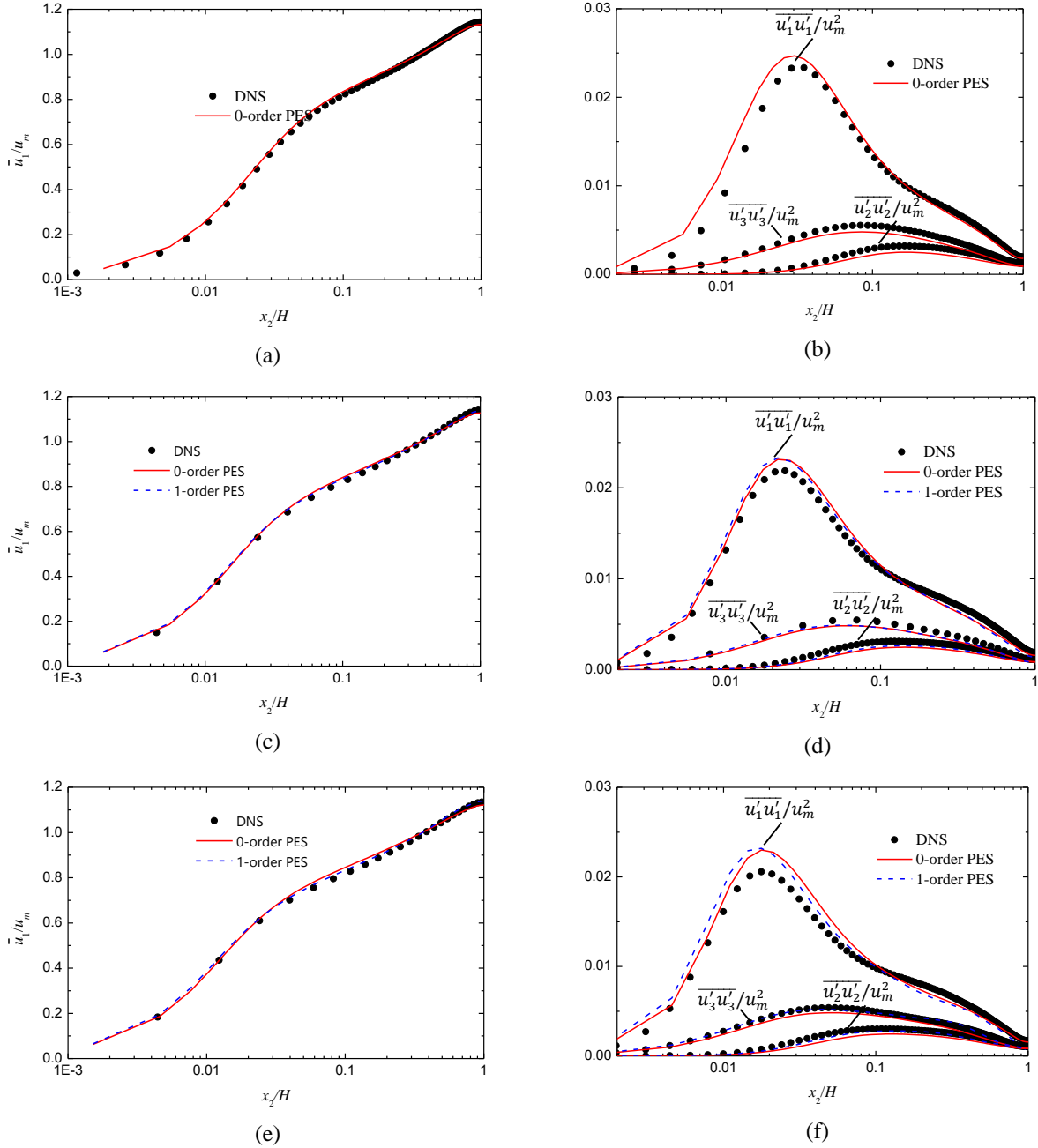

Figure 7. Distribution of the statistical results. Symbols: DNS results; Lines: PES results. (a), (c), and (e): mean velocity $\bar{u}_1$; (b), (d), and (f): Reynolds stresses $\overline{u'_i u'_j}$. (a) and (b): Re=8000; (c) and (d): Re=12000; (e) and (f): Re=16000.

### 4.3. Rough wall channel flows

The geometry for this test case is a channel with two rough walls. The rough walls are made of 2-dimensional square-shaped riblets of size $d$ which are a distance of $d$ from each other in $x_1$-direction, see Fig. 8. The height of the channel which is defined as the distance between the mid-planes of the riblets of the two side walls is $2H$. The $x_2 = 0$ plane lies at the tip of the bottom riblets.

In [31], we have calculated the same type of flows using DNS and eight RANS models, including the kε-, kω-, and RSM-family models. The comparison between the RANS and DNS results shows that the model errors of the RANS models under consideration are all higher than 20%. The kε- and RSM-family models even predicted the wrong trend with respect to the $f \sim \text{Re}$ relationship.

The DNS cases in [31] were calculated again with a smaller computational domain, which has the size $2H \times 2.1H \times H$. A DNS solution for a higher Reynolds number (Re=22400) has been added. The friction coefficients $f$ obtained from the current DNS solutions are only marginally different from the those in [31]. More computational parameters and global results for our DNS cases are shown in table 1. The Reynolds number based on the friction velocity $\text{Re}_\tau$ in table 1 is defined as

$$\text{Re}_\tau = \frac{u_\tau H}{\nu} \tag{4.5}$$

where the friction velocity $u_\tau$ for a rough wall channel flow is calculated as

$$u_\tau = \left(\nu \, \partial \langle u_1 \rangle / \partial x_2 |_{x_2=0} - \langle u_1' u_2' \rangle |_{x_2=0}\right)^{1/2} \tag{4.6}$$

The rough wall channel flows for Re=5600, 11200, 22400, 44800, and 89600 were calculated using the PES method. The maximum value of $\text{Re}_\tau$ is 6010, which is for the case Re=89600. The flow reaches the fully rough regime as Re$\geq$ 11200, the corresponding dimensionless roughness height is $d^+$=75, where $d^+$ is defined as

$$d^+ = \frac{u_\tau d}{\nu} \tag{4.7}$$

The value of $d^+$ for Re=89600 is 601.

The $f$-$\phi$ convergence study is shown in Fig. 9. It can be seen that $f$ is almost independent of $\phi$ if $\phi$ is smaller than 0.005. Thus, the CES solution for $\phi = 0.004$ is used as the reference solution (or 0-order PES solution). The 1-order and Richardson- PES solutions of $f$ are identical to the reference solution. Fig. 10 shows the 0-order PES, RANS, and DNS solutions for $f$. Obviously the PES solution has a much higher model accuracy than the RANS solutions.

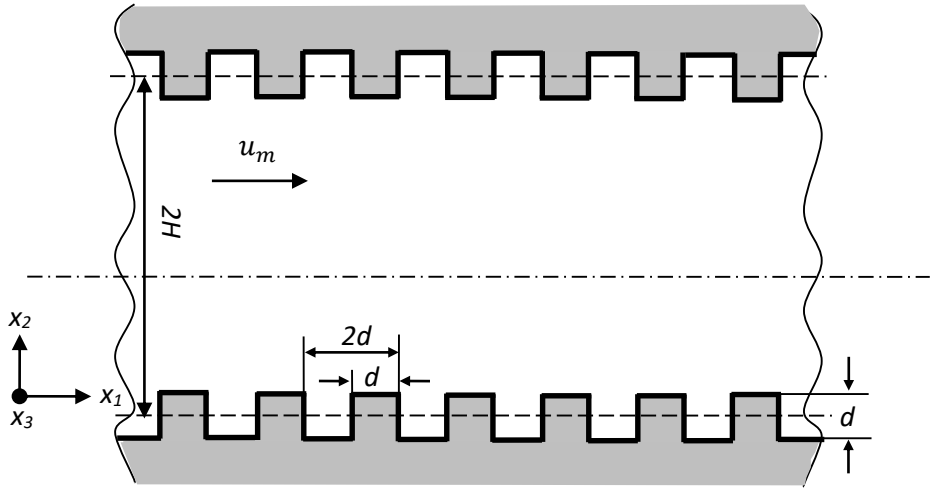

Figure 8. Schematic geometry of a plane channel with rough walls.

Table 1 Computational parameters and some global results for the DNS cases. Results obtained from DNS are marked grey.

| Re | Mesh ID | Mesh resolution | f | $Re_\tau$ |
|---|---|---|---|---|
| 2800 | A | $280 \times 220 \times 70$ | 0.046 | 206 |
| 5600 | A | $280 \times 220 \times 70$ | 0.040 | 387 |
| 11200 | B | $960 \times 326 \times 130$ | 0.036 | 732 |
| 22400 | C | $1200 \times 370 \times 170$ | 0.034 | 1485 |

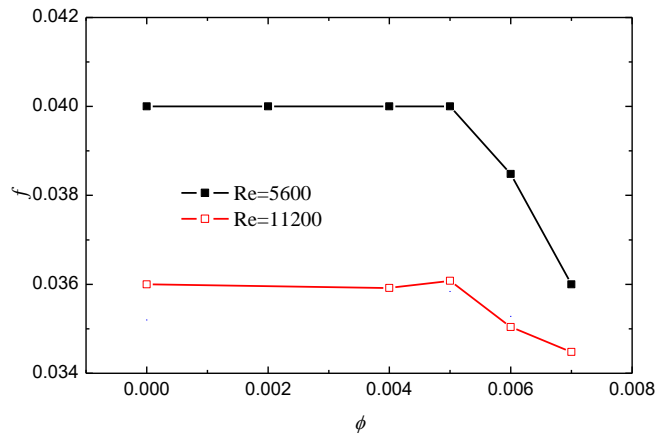

Figure 9. Friction coefficient $f$ versus weight coefficient $\phi$.

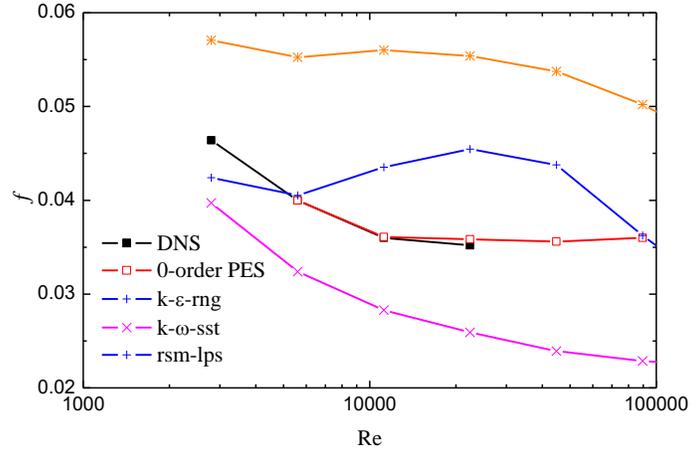

Figure 10. Friction coefficient $f$ versus weight coefficient $\phi$, the RANS solutions are taken from [31].

The instantaneous vortical structures for Re=11200 are shown in Fig. 11. The 0-order PES results are compared with the DNS results. The vortical structures are identified by the iso-surfaces with constant $Q$ values. The quantity $Q$ is the second invariant of the instantaneous velocity gradient tensor, which is defined as $-\frac{1}{2}\frac{\partial u_i}{\partial x_j}\frac{\partial u_j}{\partial x_i}$ for an incompressible flow. Hunt *et al.* [32] suggested that the vortcial structures can be identified by positive $Q$. The vortical structures are colored with the velocity magnitude. Some vortices which should populate close to the channel center have been filtered in the CES solution. However, the main features of the vortical structures have been captured.

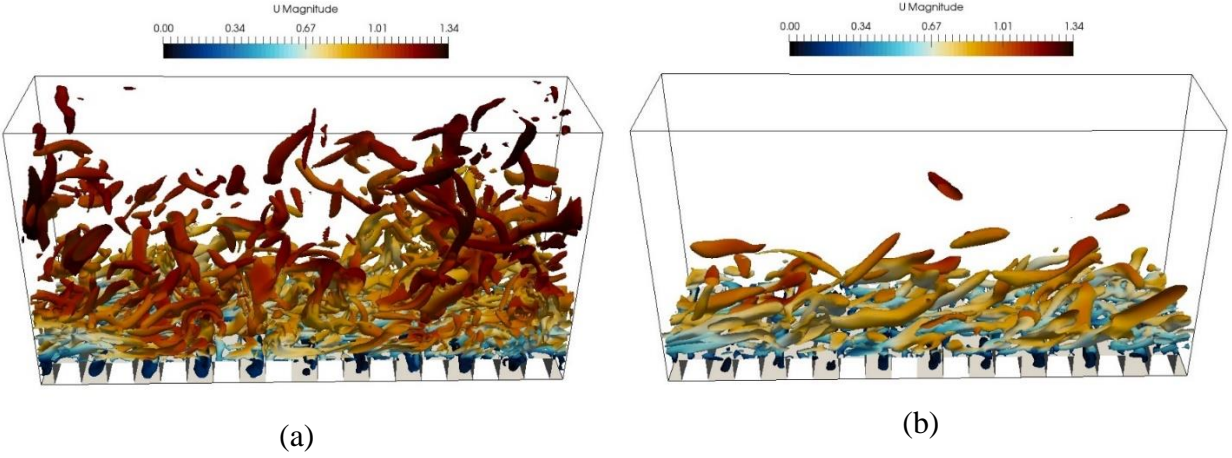

(a)             (b)

Figure 11. Snapshots of the instantaneous vortical structures in a box ($2H \times 1.05H \times 0.5H$, 1/4 flow domain); iso-surfaces $Q(H^2/u_m^2) = 5$, Re=11200. (a) DNS results; (b) 0-order PES (reference CES) results.

The statistical results including the mean velocity $\bar{u}_1$ and the Reynolds stresses $\overline{u_i' u_j'}$ for Re=5600, 11200, and 22400 are shown in Fig. 12. The 0-order PES solutions are in good accordance with the DNS results. The logarithmic region of $\bar{u}_1$ was successfully predicted. The PES results show that the logarithmic region is enlarged from $30 \leq x_2^+ \leq 160$ to $30 \leq x_2^+ \leq 800$ as the Reynolds number increases from 5600 to 22400. The normalized velocity $u_1^+ = \langle u_1 \rangle / u_\tau$ in this region can be approximated as

$$u_1^+ = \frac{1}{0.41} \ln(x_2^+) + C - \Delta u^+ \qquad (4.8)$$

where $C = 5.5$ is a constant. The roughness function $\Delta u^+$ will increase from 2.2 to 4 as Re increases from 5600 to 22400. These results are in accordance with the theories for turbulent flows over a rough wall.

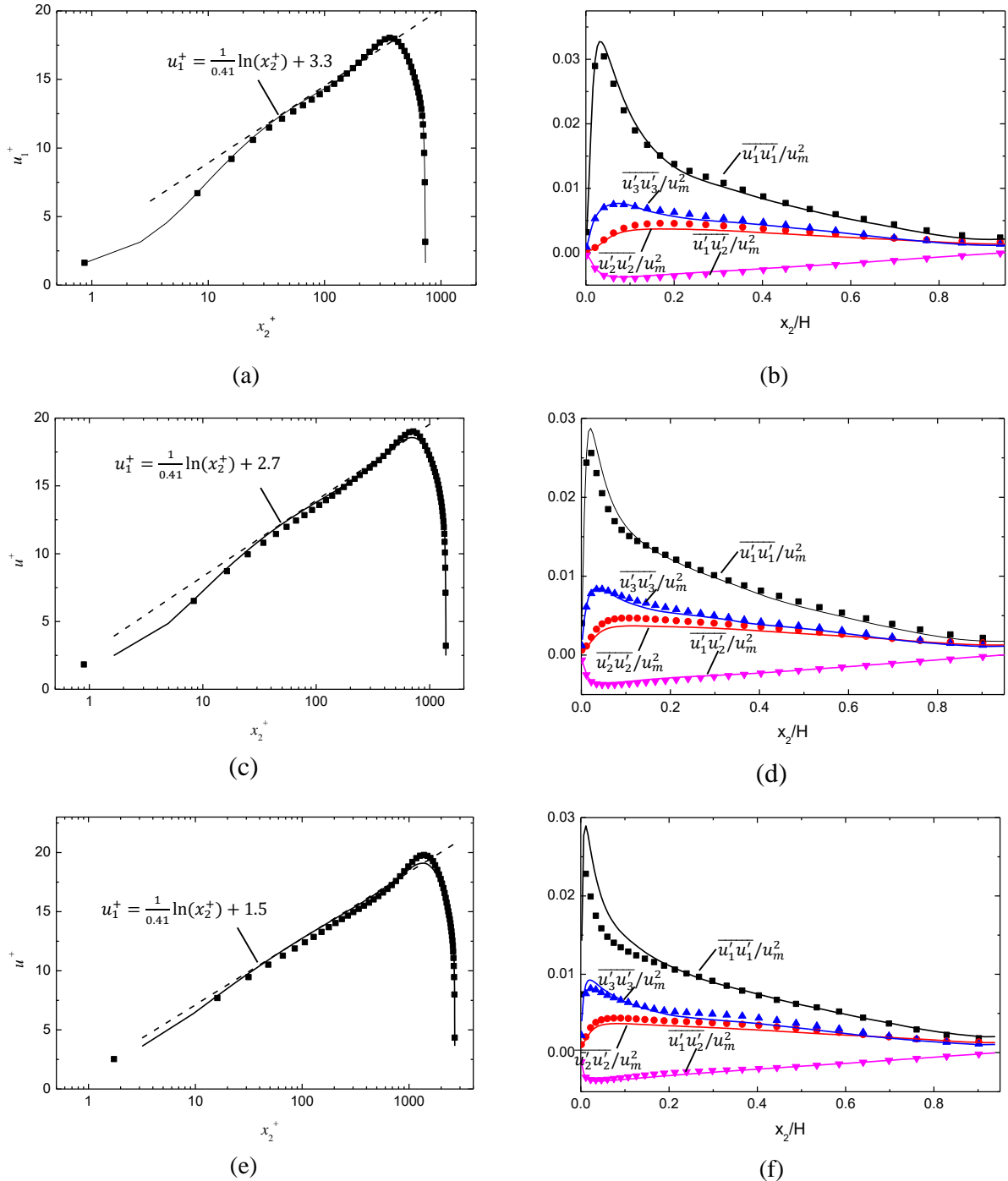

Figure 12. Distribution of the statistical results. Symbols: DNS results; Lines: 0-order PES results. (a), (c), and (e): mean velocity $\bar{u}_1$; (b), (d), and (f): Reynolds stresses $\overline{u'_i u'_j}$. (a) and (b): Re=5600; (c) and (d): Re=11200; (e) and (f): Re=22400.

4.4. Compressor blade cascade flows

The turbulent flow in a compressor cascade was calculated by using the PES method. The computational domain is a passage of the cascade made of airfoils NACA0065-009. A periodic boundary condition in the pitchwise direction was used to account for the effects of the neighboring airfoils.

In order to reduce the boundary effects, the inlet and outlet regions were extended by $1.59c$ and $2c$, where the chord length is $c = 0.15$m. The velocity and turbulence intensity profiles at the inlet were given according to the experimental data in [33]. The free stream velocity at the inlet $u_\infty$ is 40ms$^{-1}$. A divergence free synthetic eddy method [34] was used to approximate inlet velocity fluctuations. Two incidence angles $\alpha$ were accounted for in the study. They are 0° and 4°. Since the flow is at a small Mach number (smaller than 0.3), the fluid in the cascade can be assumed to be incompressible. The flows have the Reynolds number $3.82 \times 10^5$, which is based on the chord length $c$ and the free stream velocity $u_\infty$.

Tripping bands (3.0 mm wide by 0.3 mm thick, located at 6.0 mm from the leading edge) were used in the experiments of [33] in order to remove the difficulty for simulating laminar-turbulence transition. Although laminar-turbulence transition can be directly resolved from a CES solution, in order to compare with the experimental data, we still repeated our test cases with the tripping bands being accounted for. The effects of tripping bands were modeled by adding a Darcy's term $\frac{\nu}{K}\tilde{u}_i$ in the momentum equation (2). The permeability $K$ was calculated according to the Kozeny's equation [35]:

$$K = \frac{d^2 \phi^3}{\beta(1-\phi)^2} \quad (4.9)$$

According to [33], the particle diameter $d$ is $25\mu m$, the porosity $\phi$ is 0.48, the model constant $\beta$ is 180.

A body fitted mesh which concentrates near the airfoil surface and bounded walls was adopted in the study, see Fig. 1. The dimensionless mesh spacing $y_w^+$ of the first grid point near the wall is smaller than 1, thus the turbulent boundary layer can be resolved. The mesh in the region close to the cascade trailing edge is refined to capture the boundary layer separation. The standard mesh which was used in our test cases has about 16 million mesh cells, with 101173 cells in the $x_1$-$x_2$ plane and 160 cells in the spanwise ($x_3$-) direction. Fig. 13 shows the mesh in a half computational domain. A traditional LES of the same cascade flow but for only a half span used 200 million grid points, see [36]. A CES solution uses much lower computational resources.

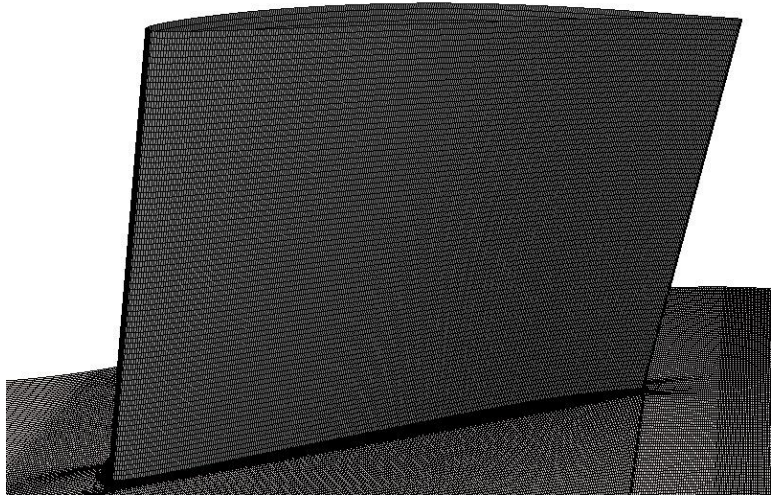

Figure 13. The mesh resolution in a half domain.

The test case for $\alpha = 0°$ with a higher mesh resolution (about 83 million cells) to study the sensitivity our CES solution to the mesh resolution. For a cascade flow, the losses of mechanical energy corresponds to the stagnation pressure coefficient $\omega_1$. The $\omega_1$-$\phi$ convergence study shows that $\omega_1$ is only marginally changed as the value of $\phi$ is increased from 0.007 to 0.009. Therefore, we use the CES solution for $\phi = 0.007$ as the reference solution. The 1-order and Richardson-PES solutions for $\omega_1$ are identical to the 0-order PES (or reference CES) solution.

The time averaged pressure coefficients $\bar{c}_p$ at the airfoil surface for $\alpha = 0°$ and $4°$ are compared with the experimental data in Fig. 14. The $x_2$- line averaged stagnation pressure loss coefficients $\omega_{13}$ at the section $x_1$=1.363$c_a$ are compared with the experimental data in Fig. 15. The 0-order PES results are in good accordance with the experimental results. Similar to the LES results in [36], our calculated $\bar{c}_p$ for $\alpha = 4°$ is also slightly higher than the experimental results. A possible reason for this discrepancy is that the inlet conditions for the simulation and those for the experiment may still have some differences.

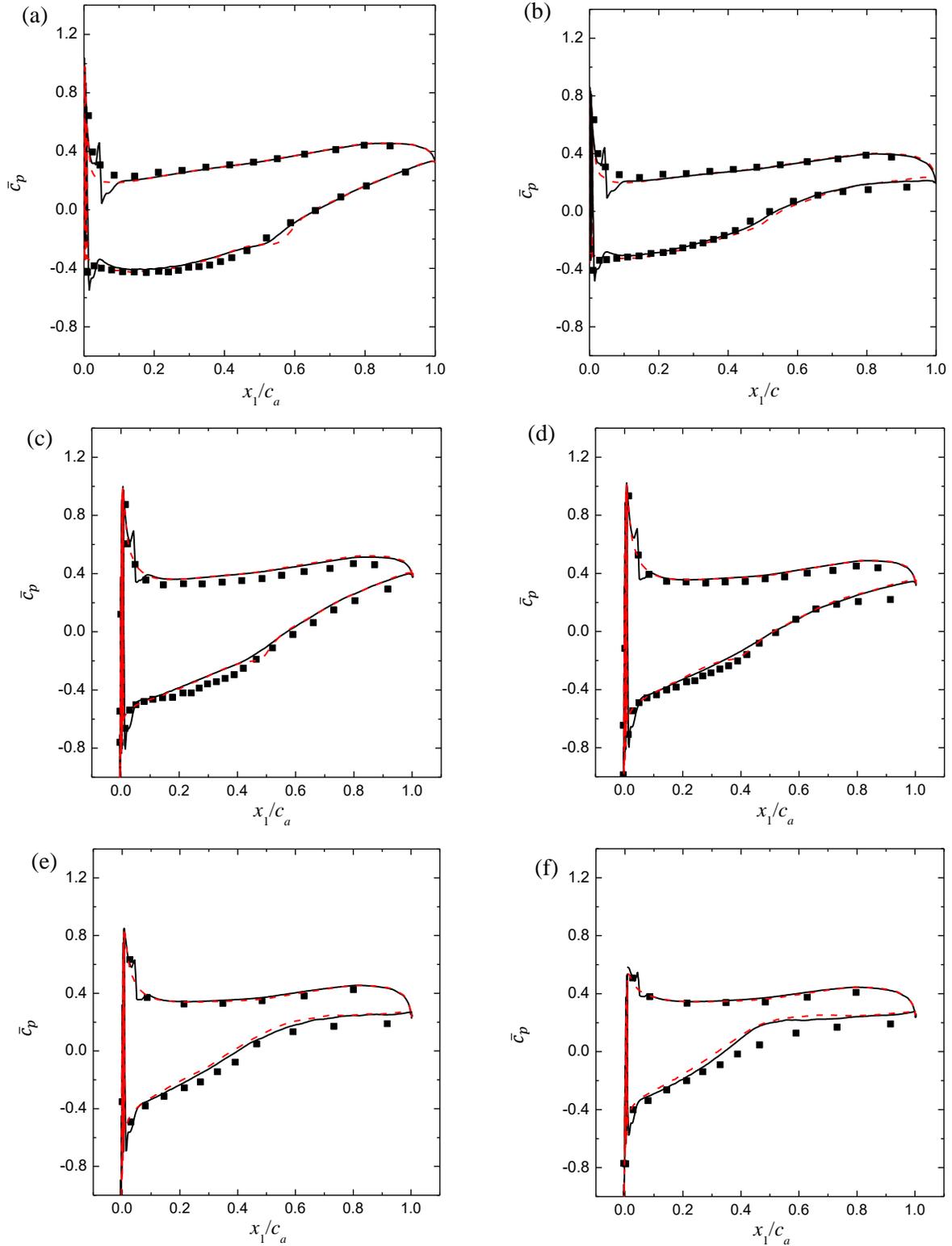

Figure 14. Time averaged pressure coefficient $\bar{c}_p$ at the airfoil surface. ■: Experimental results in [33]; —: 0-order PES solution, with tripping bands; ---: 0-order PES solution, without tripping bands. (a) $\alpha = 0°$, $x_3/h = 50\%$; (b) $\alpha = 0°$, $x_3/h = 5.4\%$; (c) $\alpha = 4°$, $x_3/h = 50\%$; (d) $\alpha = 4°$, $x_3/h = 16.2\%$; (e) $\alpha = 4°$, $x_3/h = 5.4\%$; (f) $\alpha = 4°$, $x_3/h = 1.4\%$.

A laminar-turbulence transition may occur at the airfoil surface if the tripping bands are not used. The transition leads to a jump of the time averaged pressure coefficient at the suction surface close to the mid-span, see Figs. 14(a), (c) and (d). The transition can be also identified by the distribution of the mechanical energy losses $\tilde{\varepsilon}$. Fig. 16(a) shows that $\tilde{\varepsilon}$ starts to oscillate as the laminar-turbulence transition occurs. An early transition due to the tripping bands leads to a fluctuation of $\tilde{\varepsilon}$ close to the leading edge of the airfoil, see Fig. 16(b).

Fig. 17 shows the vortical structures identified by the iso-surfaces of $Qc^2/u_\infty^2=1.4$. The iso-surfaces are colored with the velocity magnitude. The 0-order PES results show that the secondary flows near the end-walls are dominated by the horseshoe vortex, which can be decomposed as the leading edge horseshoe vortex, the pressure side leg vortex, and the suction side leg vortex. The horseshoe vortex originates at the adjunction of the endwall and the airfoil leading edge. The pressure side leg of the horseshoe vortex is driven by the pressure gradient to the adjacent suction surface. The pressure and suction side leg vortices eventually merge each other to form the passage vortex. These structures are in accordance with the experimental observations. In general, the 0-order PES solution has a similar accuracy as the LES solution in [36], while its computational cost is much lower.

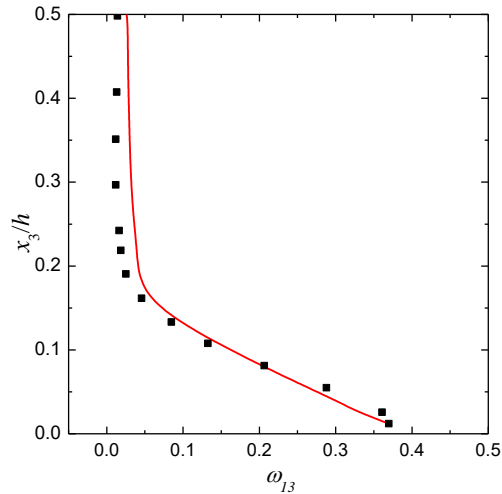

Figure 15. $x_2$- line averaged stagnation pressure loss coefficients $\omega_{13}$ at the section $0.363c_a$ downstream the airfoil trailing edge, $\alpha = 4°$, with tripping bands. ■: Experimental results in [37]; –: 0-order PES solution.

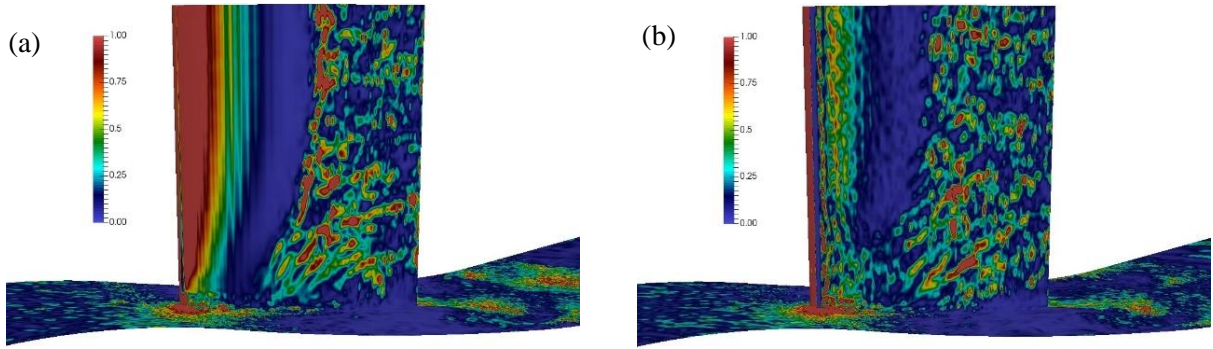

Figure 16. Instantaneous dissipation rate $\tilde{\varepsilon}c/u_\infty^3$ at the wall surfaces, facing the suction surface, $\alpha = 4°$, the 0-order PES (reference CES) results for a half span. (a) without tripping bands; (b) surface, with tripping bands.

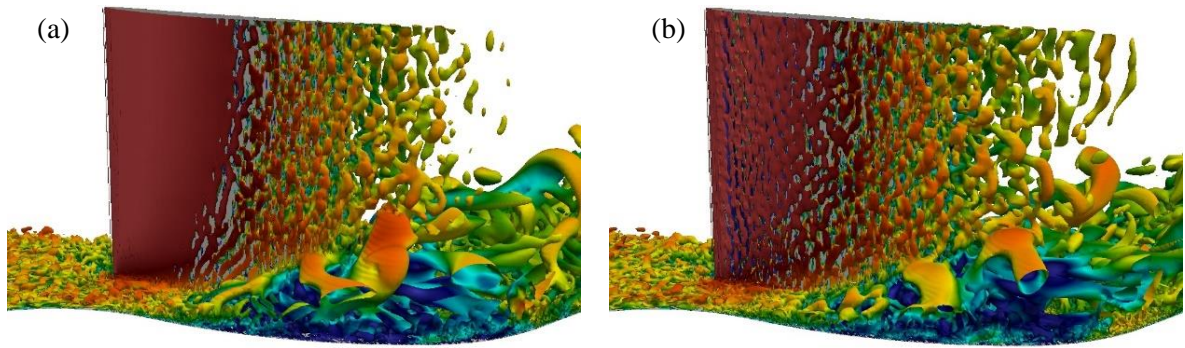

Figure 17. Turbulent structures identified by the iso-surfaces of $Qc^2/u_\infty^2 = 1.4$, facing the suction surface, the 0-order PES (referece CES) results for a half span, facing the suction surface, $\alpha = 4°$. (a) without tripping bands; (b) with tripping bands.

## 5. Conclusions

Parameter extension simulation (PES) as a mathematical method for simulating turbulent flows has been proposed. It is defined as a calculation of the turbulent flow for the desired parameters with the help of a reference solution. A controlled eddy simulation (CES) method has been developed to calculate the reference solution $R_0$. The CES method is a special type of large eddy simulation (LES) method in which an artificial force distribution $F_i$ and a weight coefficient $\phi$ are used to model part of the turbulent motions.

A modified mixing length model (ML+) has been proposed to calculate $F_i$. The reference $\phi$ value is determined through a $\phi$-convergence study. Since $\phi$ is uniformly distributed in space, the established mesh-convergence study methods (such as the Richardson extrapolation) can be used in the $\phi$-convergence study. 0-order, 1-order PES, and Richardson- PES methods have been proposed in the study.

The proposed PES/CES method has been used to simulate four types of turbulent flows to demonstrate its application. They are decaying homogeneous and isotropic turbulence, smooth wall channel flows, rough wall channel flows, and compressor blade cascade flows. For a turbulent flow at a high Reynolds number, the 0-order PES (or reference CES) solution has a similar accuracy as a traditional LES method (see Fig. 5), while it is much more accurate than RANS methods (see Fig. 10). It can capture all the important features for the compressor cascade flow including the laminar-turbulence transition and the horseshoe vortices, see Figs. 16 and 17. The statistical results are in good accordance with the experimental data in [33] and [37]. The model accuracy can be further improved by adopting a higher order PES method, see Figs. 2 and 5.

The computational costs for a PES solution is much lower than those for a traditional LES solution at a high Reynolds number. When the same mesh resolution is adopted, the numerical error $\delta_{num}$ for a 0-order PES solution is lower than that for a traditional LES solution (see Fig. 6). Therefore, to achieve the same numerical accuracy, we expect that a PES solution requires fewer mesh cells than a traditional LES solution. PES of a compressor cascade flow needs only about 1/25 of the cells for a traditional LES study, see [36]. Our test cases show that the PES method is particularly suitable for turbulent flows with a high Reynolds number in a complex geometry.

## Acknowledgments

The authors gratefully acknowledge the support of this study by the DFG-Heisenberg program (299562371) and the North-German Supercomputing Alliance (HLRN). The acknowledgement is also given to Prof. H. Herwig of Hamburg University of Technology for the helpful discussion with him.